\documentclass[final,3p,times,twocolumn]{elsarticle}
\usepackage{amssymb}
\usepackage{soul}
\usepackage{amsmath}
\usepackage{float}
\usepackage{xcolor}
\biboptions{sort&compress}
\usepackage{color}
\definecolor{RED}{RGB}{156,78,90}

\journal{Journal of Colloid and Interface Science}
\definecolor{B}{RGB}{52,78,200}

\begin{document}
\begin{sloppypar}
\begin{frontmatter}

\title{Ionic environment-modulated nucleation and stability of multiscale nanodomains in surfactant-free microemulsions}

\author[1]{Yawen Gao}
\address[1]{New Cornerstone Science Laboratory, Center for Combustion Energy, Key Laboratory for Thermal Science and Power Engineering of Ministry of Education, Department of Energy and Power Engineering, Tsinghua University, Beijing 100084, China}

\author[1]{Changsheng Chen}

\author[2]{Mingbo Li\corref{cor1}}
\ead{mingboli@sjtu.edu.cn}
\cortext[cor1]{Corresponding author}
\address[2]{Key Laboratory of Hydrodynamics (Ministry of Education), School of Ocean and Civil Engineering, Shanghai Jiao Tong University, Shanghai 200240, China}

\author[1,3]{Chao Sun\corref{cor1}}
\ead{chaosun@tsinghua.edu.cn}
\address[3]{Department of Engineering Mechanics, School of Aerospace Engineering, Tsinghua University, Beijing 100084, China}

\begin{abstract}
Nano-clustering occurs in the monophasic ``pre-Ouzo" region of ternary liquid mixtures without the use of surfactants. This study is proposed to elucidate the nucleation and stability of multiscale nanodomains in a surfactant-free microemulsion (SFME) system composed of trans-anethol, ethanol and water, tuned by aqueous ionic environment. We examined direct- and reverse-SFME structures with different compositions performed by dynamic light scattering and nanoparticle tracking analysis. Our findings demonstrate that extreme pH and ionic strength modulation significantly affect the nucleation and growth of nanodomains, resulting in larger sizes and lower number density of mesoscopic droplets in O/W structuring, whereas the reverse aggregates (W/O structuring) remains stable. Our findings clarify that Ostwald ripening is the primary mechanism to drive the droplet growth. Both theoretical calculation and experiment results match well and reveal that droplet ripening is significantly affected by extreme acidic (pH\textless3) and alkaline (pH\textgreater10) conditions. Electrostatic repulsion in a neutral ionic environment can prevent the coalescence of droplets induced by collision. This research provides an insight into the behavior of multiscale nanodomains in SFME under extensive pH control for applications in material synthesis, drug solubilization and gel preparation.
\end{abstract}


\end{frontmatter}

\section{Introduction}

Surfactant-free microemulsions (SFMEs)~\cite{xu2018surfactant, yan2017modular, yan2021nanoprecipitation}, also known as detergentless microemulsions~\cite{ma2025thermodynamic}, pre-Ouzo~\cite{klossek2012structure}, micellar-like structural fluctuations~\cite{chen2024unraveling}, mesoscale solubilization~\cite{subramanian2013mesoscale, robertson2016mesoscale}, or ultraflexible microemulsions~\cite{prevost2016small}, have received a lot of attention recently, offering a more environmentally friendly and potentially safer alternative in colloidal science~\cite{zhang2020temperature}. The stabilization of SFMEs can be achieved through the use of amphiphilic molecules or by exploiting the inherent properties of the components~\cite{hsu2022observation}. In the case of "pre-ouzo", this region is the single-phase domain just before macroscopic phase separation (the classic ouzo effect) sets in. It represents a delicate metastable equilibrium in which surfactant-free nanoscopic oil-water clusters are thermodynamically stabilised by hydrotropic molecules. Weak surfactant-free aggregates already exist, but the system remains optically clear until a critical composition is crossed~\cite{xu2017synthesis, zemb2016explain}. Based on variation on the mixed ratio, microemulsion structures include oil-in-water (O/W), bicontinuous (BC) and water-in-oil (W/O) structures, which can be clearly identified and distinguished in the diagram~\cite{han2022formation,liu2018surfactant}.

Amphiphilic molecules, also called hydrotrope molecules, usually consist of smaller nonpolar tails and cannot self-assemble to form micelles in bulk or films at the water/oil interface~\cite{eastoe2011action}. They enhance dynamic molecular clustering through hydrogen bonding with water molecules and exhibit strong similarities to cosolvents at both the nanoscale and macroscopic thermodynamic levels~\cite{kunz2016hydrotropes}. Amphiphilic molecules preferentially adsorb (or enrich) at the water/oil interface, leading to a significant decrease in interfacial tension and forming an effective interfacial layer that separates the polar water and apolar oil domains~\cite{marvcelja2011hydration, donaldson2015developing, lopian2016morphologies, schottl2019combined, kuchierskaya2021interfacial}. However, unlike traditional surfactants, hydrotrope molecules do not form a ``rigid" ordered monolayer at the interface. As a result, the curvature energy of the interface layer do not significantly contribute to the free energy of the system~\cite{zemb2016explain}. The behavior resembles more the weak specific attraction of ions to the interface~\cite{jungwirth2006specific}.

In SFMEs, the well-organized nanodomains have prompted vigorous debate regarding their origin and thermodynamic stability. These nanodomains are well-defined and discrete entities that can exist from hours to years~\cite{zhou2021co2, xu2013surfactant, hou2016surfactant, han2022formation}. The size of SFME ranges from approximately $\sim$1 nm, where it exists as molecular aggregates, to $\sim$200 nm, where it forms mesoscopic nanodomains~\cite{diat2013octanol, schoettl2014emergence, schottl2016aggregation}. A theoretical framework, considering the balance between hydration force and entropy, has been proposed and suggested to be a basic mechanism for the spontaneous formation and stability of the nanostructures~\cite{zemb2016explain}. The hydration interactions can effectively overcome entropy, resulting in the insertion of net repulsive forces between adjacent aggregates. From other perspectives, the mesoscale ordering of weak aggregates may be attributed to the attractive interactions at the molecular level~\cite{qiao2015molecular} or competition between short-range and long-range interactions~\cite{sweatman2014cluster, sweatman2019giant}.  

Although remarkable progress has been made in both experimental analysis and theoretical calculation on SFMEs, understanding on their dynamic growth remains limited. Following the previous research pathway, there are many unresolved issues regarding SFMEs that need further clarification~\cite{qi2023surfactant, gradzielski2021using, han2022formation}. The stability of nanodomains is highly sensitive to external stimulis, particular the ionic environment~\cite{li2022spontaneously, li2023thermal}. Specifically, strong ionic strength can affect both the solubility of oil in water and ionization states of the amphiphilic molecules ~\cite{salabat2024ionic, schottl2018salt}. However, the impact of such extreme ionic conditions on the formation of nanodomains in terms of size distribution, growth rate and the final morphology over time has not been fully understood. For example, how does the ionic environment manipulate the structure of nanodomains? What is the extent of ion absorption at the oil-water interface, thus causing the competitive adsorption with amphiphilic molecules? Furthermore, what is the threshold salt concentration that determines droplet stability? Addressing these questions is crucial for applications in the pharmaceutical and cosmetic industries, as a properly optimized aqueous phase is key to ensuring the solubilization, biocompatibility, and efficacy of ingredients, as well as enhancing drug delivery processes \cite{lohse2020physicochemical}.

In this work, an SFME is presented, composed of trans-anethole, ethanol and water, with trans-anethole and ethanol acting as the oil phase and amphipathic solvent, respectively. Both direct-SFME and reverse-SFME structurings within the monophasic region were selected as the primary focus. Characterization of the nanodomains, including size, number density, Zeta potential and dielectric constant, has been conducted under varing ionic conditions. Additionally, the dynamic growth of nanodroplets was monitored over a 6-hour period following SFME preparation, supported by theoretical calculations. This study provides the first detailed investigation of the role of extreme pH and ionic strength in modulating the nucleation and stability of nanodomains in SFMEs, which is essential for developing a reliable and controllable approach to nanomaterial synthesis. Furthermore, this work elucidates a fundamental mechanism of eco-friendly SFMEs in the realms of drug delivery, cosmetics and environmental remediation.

\section{Experiments}

\subsection{Chemicals and Materials}

All mixture were prepared using trans-Anethol ($\rm C_{10}H_{12}O$, Purity:\textgreater 99\%, CAS: 11-78-70, molar mass: 148.2 g/mol), which was purchased from Sigma-Aldrich. Ethanol (99.9\%, CAS: 25-84-49) was obtained from J\&K Scientiﬁc (China). Both reagents were used as received. Ultrapure water was taken from a Milli-Q purification water system (Merck, Germany) with a resistivity of 18.2 M$\Omega$ $\cdot$cm and a natural pH of 6.54 at 25 $^\circ$C. To control the aqueous environment, pH regulators, including sodium hydroxide solution (NaOH, Purity: 99.5\% AR, Sigma Aldrich, Germany) for alkalinity and hydrogen chloride solution (HCl, 37\% AR, Tongguang Chemicals, China) for acidity, were employed. 

\subsection{Preparation of surfactant-free micro-emulsion}

\begin{figure}[!t]
\centering
\includegraphics[width=0.47\textwidth]{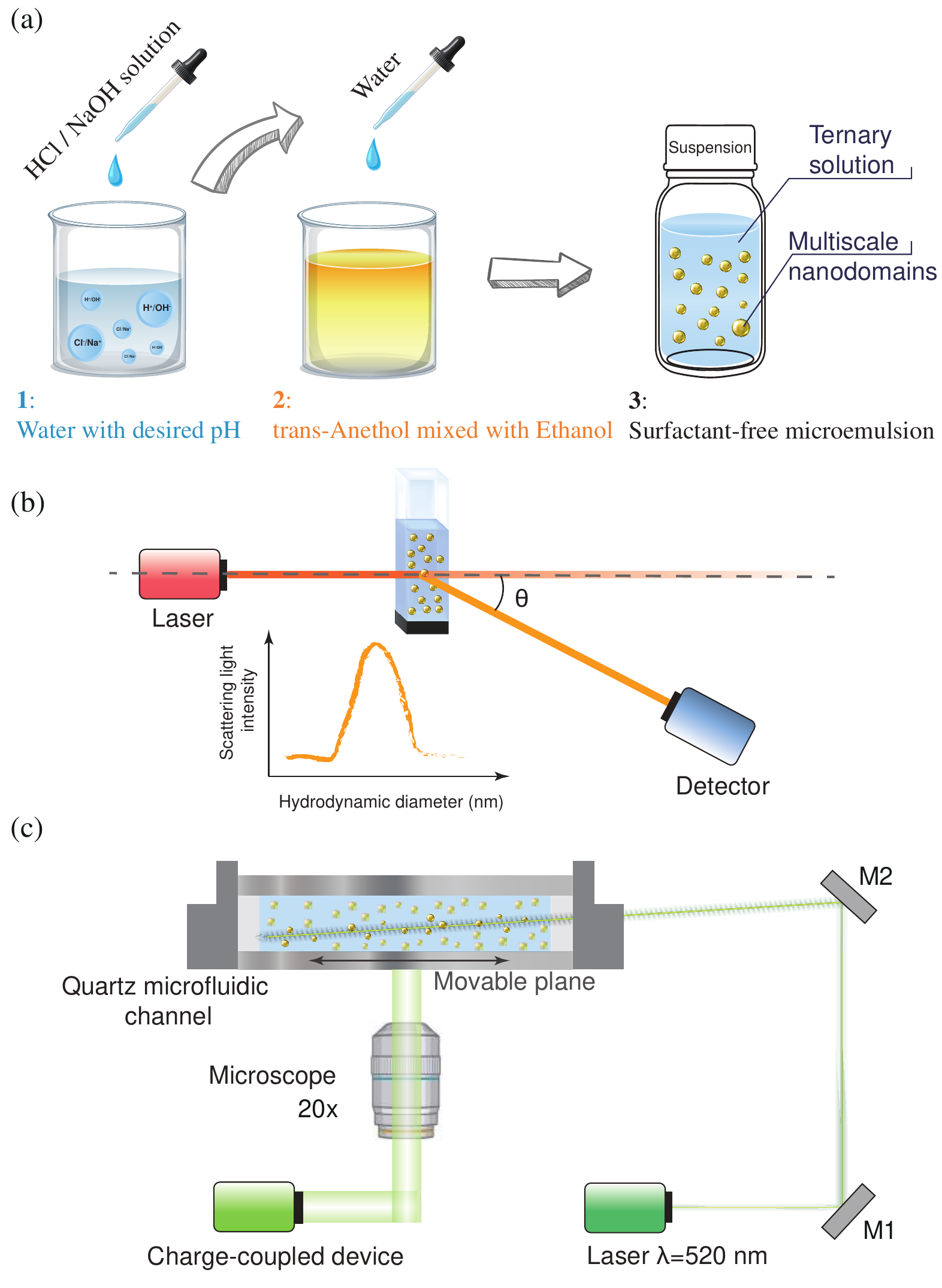}
\caption{(a) Preparation procedure of surfactant-free microemulsion. (b) Protocol of dynamic light scattering measurement on droplet size for nanodomains. (c) Schematic of the self-made nano-tracking analysis (NTA) system.} 
\label{preparation}
\end{figure}

The procedure for preparation of SFME is illustrated in Figure \ref{preparation}(a). Initially, before any procedure for sample preparation, all glassware was cleaned by placing in the ultrasonication bath for 5 min, immersing in the ethanol (99\%, AR, Titan, China) and milli-Q water. The range of pH values applied in this study is from 2 to 12. The pH was adjusted by the addition of either hydrogen chloride or sodium hydroxide solution to pure water, checking by a pH meter (pH-electrode InLab Expert Pro-ISM, METTLER TOLEDO, Switzerland) with precise pH readings. The purification of adjusted water was examined by dynamic light scattering (DLS) measurement, showing no any detected signals for dust and particles inside. Concurrently, trans-anethol at a volume fraction of $\rm \phi_o$ was mixed with certified ethanol ($\rm \phi_e$). Subsequently, water ($\rm \phi_w$) with a precisely controlled pH was slowly added into homogeneous one-phase mixture to avoid the intense turbulent mixing. The sample was formed without any addition of surfactant and immediately became whitening after water was brought into contact with oil-ethanol mixture. After mixing, the sample was placed in a mixer (Vortex-5, Kylin-Bell, China) and homogenized for 3 min to achieve a stable micro-emulsion. Then the droplet growth was immediately characterized by DLS and nanotracking tracking analysis (NTA) measurements.

The trans-anethol/ethanol/water system was chosen in this study due to its well-characterized Ouzo effect, which provides an ideal model for investigating the formation and stability of self-organized nanostructures in ternary systems. The total volume of the final product was consistently maintained at 20 mL, and the volumetric ratio of the three components satisfied an equation: $\rm \phi_o$+$\rm \phi_e$+$\rm \phi_w$=100\%. To simplify the nomenclature of SFME, for example, we used O5E75W20 to represent the sample with $\rm \phi_o$, $\rm \phi_e$ and $\rm \phi_w$ of 5\%, 75\% and 20\% in volumetric ratio, respectively. In this study, SFME samples with compositions O5E75W20, O15E75W10, and O40E55W5 were prepared across a pH range of 2–12, and an additional sample with a composition of O15E80W5 was prepared to examine the effect of salt. 

\subsection{Characterization of multiscale nanodomains}

Dynamic light scattering (DLS, ZEN3700 Zetasizer NanoZSE, Malvern Instruments, U.K.) was utilized to determine the size distribution of droplet ranging from 0.4 nm to 1 $\mu$m. The thermal fluctuations of colloidal particles or nanodroplets in the solution generate a fluctuating scattered intensity signal through a photo counter placed at a fixed scattering angle of $\theta$. The fluctuation is derived from the Brownian movement of the scattered substances. According to the Stokes-Einstein relation and field-correlation function, the DLS instrument provides the intensity of scattered light along with the hydrodynamic diameter of substances, as shown in Figure \ref{preparation}(b). The standard measurement protocol was as follows: 1 mL sample was placed into a quartz cuvette and allowed to stabilize for 2 min at 25 $\rm ^\circ C$. A scattering angle of 173 $^\circ$ was then automatically carried out. The refractive index and the dielectric constant of ternary solutions were measured in advanced, as these parameters are essential inputs for DLS measurement. To ensure the accuracy and reproducibility, two individual measurements consisting of twelve runs were performed. 

Nanotracking analysis (NTA) was employed to characterize the number density of nanodroplets suspended in the bulk \cite{gao2025characterization}. The measuring system is assembled with an incident laser light with a wavelength of 520 nm, a quartz cell with a matched configuration, an optical microscope (IX73, Olympus, Japan) equipped with a $\times$20 magnification objective, and a charge-coupled device (CCD, xiD, XIMEA, Germany). The principle of NTA measurement is shown in Figure \ref{preparation}(c). The visible area for observation irradiated by the laser beam in the quartz microfluidic chamber is 440 $\mu$m$ \times $200 $\mu$m with a height of 5.6 $\mu$m. The output signal was transmitted and processed using a XIMEA software on a desktop. The CCD sensor captured images with a full frame resolution of 1936 $\times$ 1456 pixels, with each pixel representing an actual size of 227 nm. For each analysis, twelve random images were selected and captured. The bright spots within the laser pathway were automatically quantified using ImageJ/Fiji software. The unit of the number density for nano-droplets is the number of spots counted per unit volume.

\subsection{Measurements of Zeta potential, refractive index, dielectric constant and viscosity}

The Zeta potential of the nanostructures was determined using a Zetasizer NanoZSE (Malvern Instruments, UK) equipped with a U-shaped capillary cell. The mean value of Zeta-potential was obtained from six measurements conducted at ambient pressure and room temperature (25 $^\circ$C). The number of refractive index was measured immediately after sample preparation using a refractometer (GR30, Shanghai Zhuoguang Instrument Technology, China). To ensure reliability, each measurement was performed in triplicate. More specific data about the refractive index and viscosity of samples are provided in our previous work~\cite{li2022spontaneously}. Additionally, dielectric constant and the dynamic viscosity of mixed sample, the crucial inputs for the Zeta potential detection, were assessed by a liquid dielectric constant tester (GCSTD-F, Guance. Ltd, China) and a rheometer (Discovery Hybrid Rheometer, TA Instruments, USA), respectively. The corresponding results are shown in Table S1 in Supplementary Material.

\section{Results and discussion}

\subsection{Ternary phase diagram}

\begin{figure}[!t]
\centering
\includegraphics[width=0.47\textwidth]{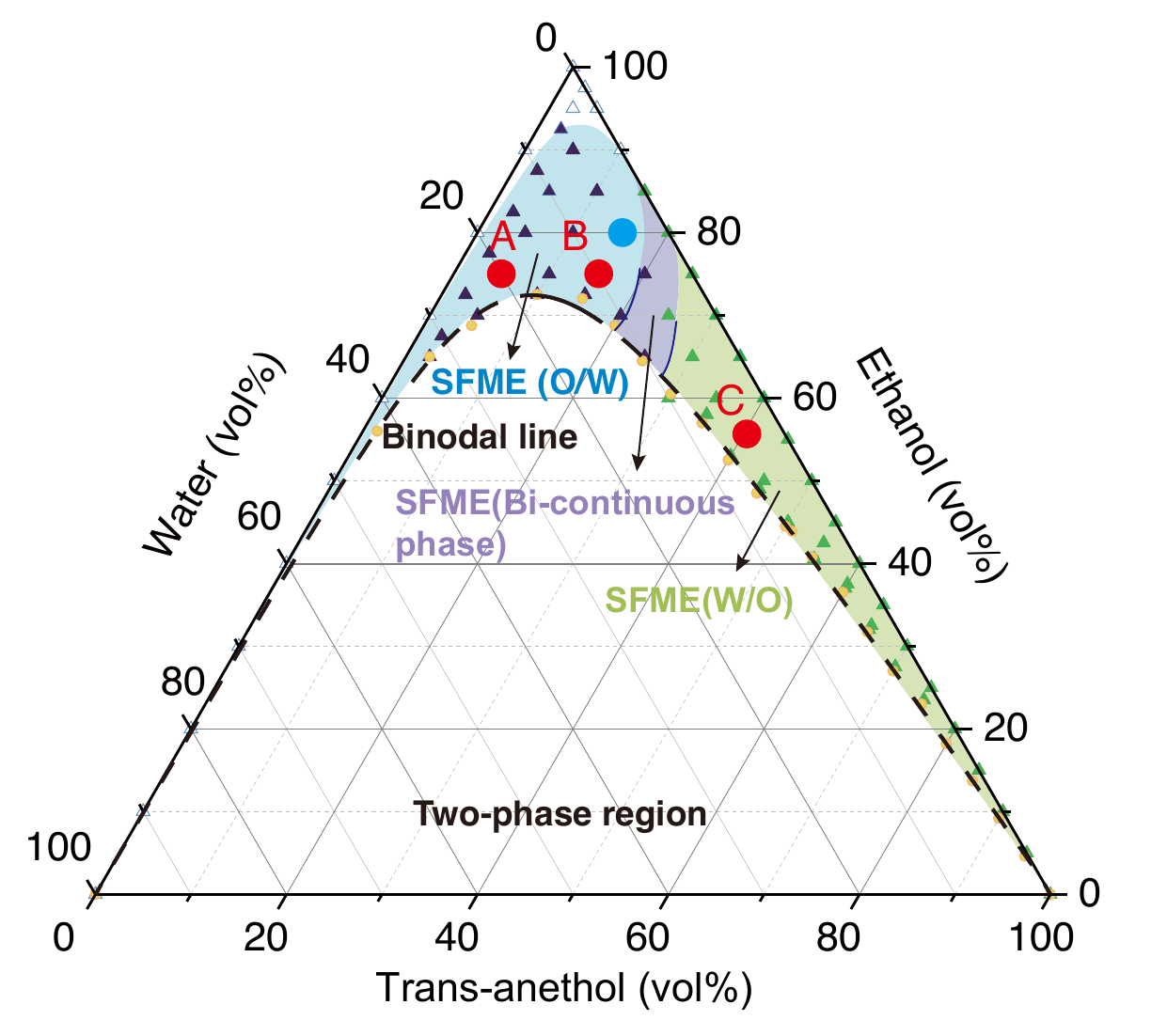}
\caption{Ternary phase diagram of trans-anethol/ethanol/water in volume fractions. It is obtained at natural pH and ambient conditions (25$^\circ$ C). Boundaries of O/W structure (blue), Bi-continuous phase (purple) and W/O structure (green) have been defined by conductivity measurements.} 
\label{FIG2}
\end{figure}

The ternary phase diagram of the trans-anethol/ethanol/water system is provided in Figure~\ref{FIG2} for clarity and ease of discussion. In this system, ethanol serves a dual role as a hydrotrope, exhibiting complete miscibility with both trans-anethol and water, which are otherwise immiscible. The phase diagram features a binodal line, depicted as a dashed curve, that defines the phase separation boundary (miscibility gap). The precise position of the binodal line in the diagram, determined using conventional titration methods, is influenced by various factors such as the component mixing ratio, pH, and temperature as reported in previous studies~\cite{blahnik2023microemulsion,vratsanos2023ouzo, li2023thermal}.

The diagram is divided into two distinct regions: a thermodynamically stable, isotropic and pseudo-monophasic region in the upper part, and a biphasic region, characterized by turbid solutions, which occupies the lower part of the diagram. Within the monophasic region, purely molecular solutions are confined to a small subset of the corners, whereas the remaining area consists of pseudo-monophasic solutions with multi-scale nanostructures. On the water-rich side, the trans-anethol component primarily exists as oil-loaded aggregates with dimensions around 1 nm and mesoscopic droplets approximately 100 nm in size. This region corresponds to the formation of oil-in-water (O/W) structure with microemulsion characteristics (SFME), depicted as the blue zone. Conversely, in the oil-rich corner, decreasing the ethanol concentration results in a gradual transition from O/W SFME to more stable ``true" W/O SFME, represented by the green zone. The boundaries of O/W structure, bi-continuous phase and W/O structure were determined by the conductivity measurements \cite{lu2024fabrication, zhang2025surfactant, prevost2021spontaneous, salabat2024ionic}. The detailed information is provided in Figure S1 of Supplementary Material. In this study, the specific composition, represented by three red circles and one blue circle on the diagram, were selected to examine the influence of pH and salty composition in the aqueous environment on the nucleation and stability of nanostructuring. 

\subsection{Dielectric properties of ternary solutions}

The dielectric constant (relative permittivity) is a fundamental material property that describes how a substance responds to an external electric field. In the context of solvent-solute interactions, the dielectric constant is crucial for understanding how well a solvent can solvate solute molecules and dissociate ions. A higher dielectric constant indicates that the solvent can better separate and stabilize charged species, such as ions, by reducing the electrostatic forces between them. This polarization capability also influences the structure of the solution and governs various physicochemical properties, such as viscosity, conductivity, and even chemical reactivity. Specifically, pH regulation plays a crucial role in modifying the ionic composition of the aqueous phase, thereby influencing the dielectric constant. By adjusting the pH from neutral to extreme values, the concentration of ions or electrolytes in the pre-prepared water phase increases, which enhances the dielectric response of the mixed emulsion.

\begin{figure}
\centering
\includegraphics[width = 0.46\textwidth]{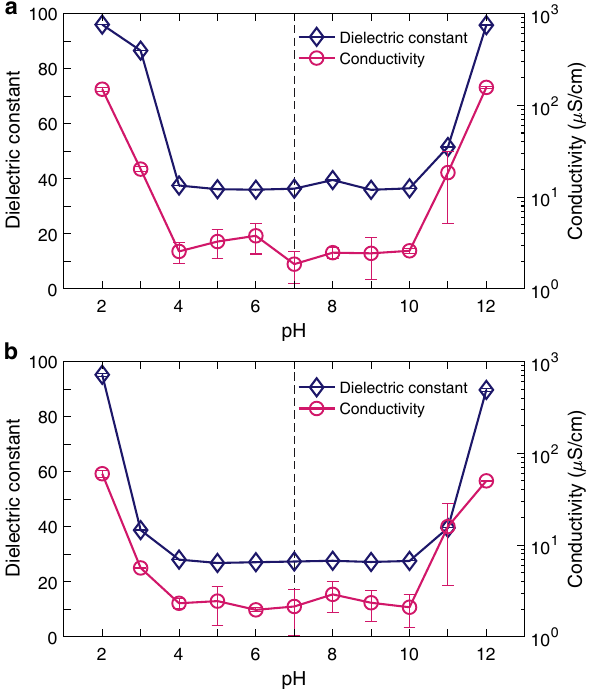}
\caption{Experimental dielectric constant and conductivity of ternary solution as a function of pH. (a) Sample A with composition of 5 vol$\%$ trans-anothel, 75 vol$\%$ ethanol and 20 vol$\%$ water. (b) Sample B with composition of 15 vol$\%$ trans-anothel, 75 vol$\%$ ethanol and 10 vol$\%$ water.} 
\label{dielectric}
\end{figure}

In Figure \ref{dielectric}(a) and (b), the dielectric constants and conductivities of two ternary solutions are plotted as a function of pH for two compositions. Sample A consists of 5 vol$\%$ trans-anethol, 75 vol$\%$ ethanol, and 20 vol$\%$ water, while Sample B contains 15 vol$\%$ trans-anethol, 75 vol$\%$ ethanol, and 10 vol$\%$ water. Both the dielectric constant and the conductivity show a clear U-shaped response to pH, with a symmetrical trend centered around pH 7. This symmetry suggests that the influence of the type of ions is considerably insignificant, and the observed behavior is predominantly governed by the concentration of ions in solution.

\begin{figure*}
\centering
\includegraphics[width = 0.93\textwidth]{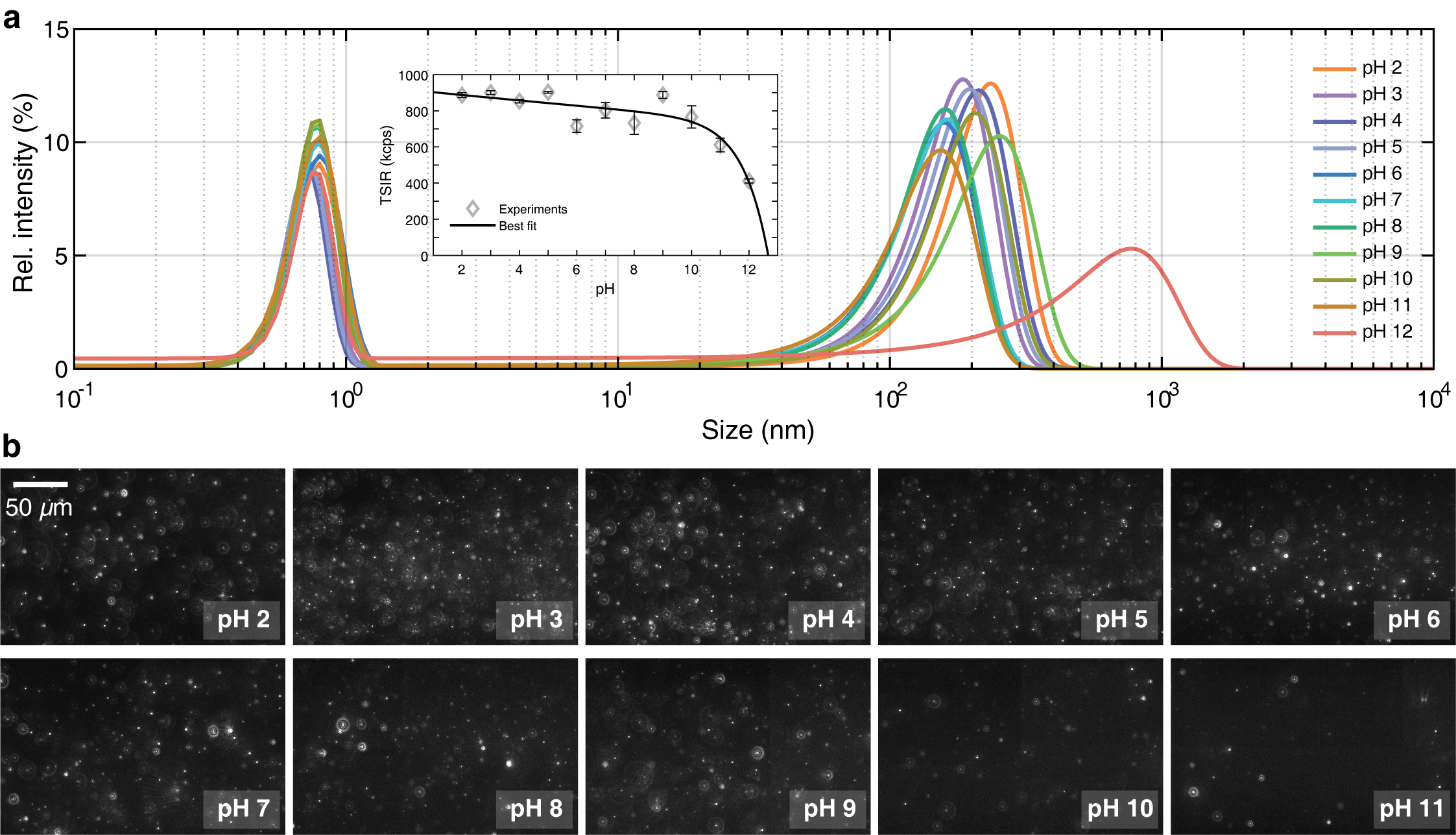}
\caption{(a) Size distribution of the multiscale nanostructures under various pH. Controllable pH is achieved by adjusting the pH of water prior to the dilution. The inset resents the dependence of the total scattering intensity rate of the mesoscale inhomogeneities (count rate, kcps) on pH. (b) A series of snapshots capturing the mesoscopic droplets under the pH range. Measurements are performed based on sample B with composition of 15 vol$\%$ trans-anothel, 75 vol$\%$ ethanol and 10 vol$\%$ water.} 
\label{O15_dls}
\end{figure*}

Experimental results indicate that both the dielectric constant and conductivity remain relatively stable within the pH range of 4 to 10 for these two samples, where the difference are primarily attributed to the composition of the solution rather than the ionic effects. For Sample A, the dielectric constant is about 40, and for Sample B, which has a reduced water content, the dielectric constant drops to approximately 30. Interestingly, when the pH is shifted beyond the neutral range either below 4 or above 10 the dielectric constant and conductivity increase dramatically. In extreme cases, the dielectric constant even approaches dielectric saturation, where the solution becomes highly polarizable and dielectric constant reaches a plateau. According to the work reported by Rambhau et al. \cite{rambhau1992influence}, there exists an inverse relationship between the interfacial area and dielectric constant in oil-in-water (O/W) structure. In other words, a mid-range pH tends to form larger interfacial areas, which are linked to the formation of smaller droplets. The extreme pH promotes the formation of larger nanodroplets, potentially contributing to an increased polarization and high dielectric response.

The composition of the ternary solution and the presence of electrolytes are the two key factors that primarily determine the dielectric constant of these solutions. The introduction of electrolytes, in particular, complicates this relationship due to the solvation structures formed around ions in the solvent. On one hand, the cations and anions from the electrolytes can become well-separated by the solvent, leading to increased ionic dissociation. This enhances the overall polarizability of the system, as the ions become more susceptible to alignment under an external electric field. The result is that electrolytes contribute significantly to the dielectric response, especially in the static dielectric constant measurements, where the system is assumed to be at equilibrium with respect to its electric polarization.

\subsection{Effect of pH on the nucleation of multiscale nanodomains}

\begin{figure}[!t]
\centering
\includegraphics[width = 0.45\textwidth]{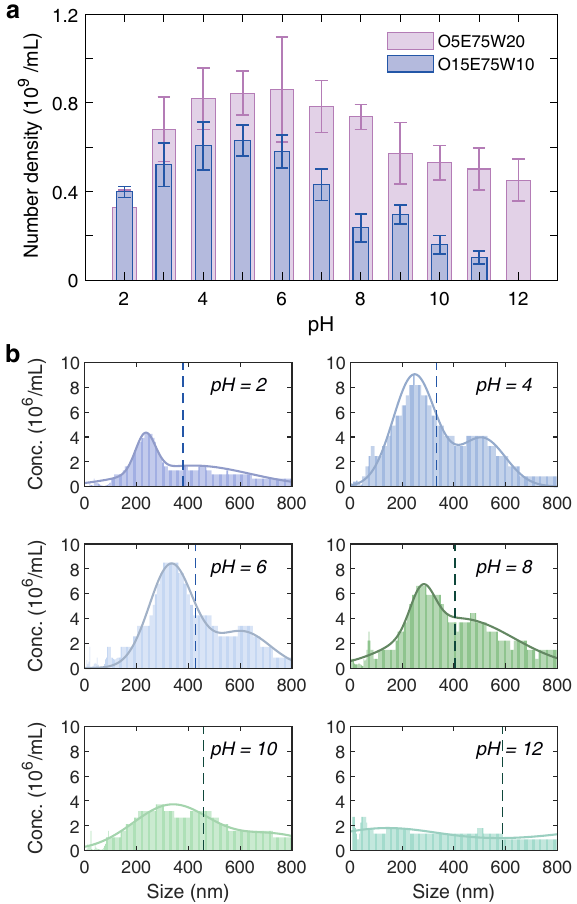}
\caption{(a) Mean number density as a function of the pH for the two samples (Sample A with composition of 5 vol$\%$ trans-anothel, 75 vol$\%$ ethanol and 20 vol$\%$ water; Sample B with composition of 15 vol$\%$ trans-anothel, 75 vol$\%$ ethanol and 10 vol$\%$ water). (b) Size-concentration distribution histogram for various pH of Sample A. Dashed lines mark the diameter of mesoscopic droplets ($D_{50}$).} 
\label{O5_dls}
\end{figure}

Surfactant-free microemulsions were successfully formed by systematically varying the concentrations of hydrogen chloride and sodium hydroxide, which served to modulate the electrolyte content in the system. The electrolyte concentration, in terms of pH, was adjusted within a range of $10^{-7}$ and $10^{-2}$ mol/L, corresponding to pH values between 2 and 12 in the aqueous phase before it reached equilibrium with the trans-anethol and ethanol components. This precise control over pH allowed for a detailed investigation into the effects of electrolyte's ionic strength on the phase behavior of an SFME system composed of 15\% trans-anethol, 75\% ethanol, and 10\% water, as illustrated in Figure~\ref{O15_dls}. DLS measurements were conducted at a controlled temperature of 25$^\circ$C to examine the evolution of the size distribution under varying pH conditions.

As previously reported~\cite{li2022spontaneously}, the size distribution spectra obtained from DLS reveal the presence of two distinct peaks, indicative of the coexistence of two types of structural inhomogeneities: molecular-scale aggregates (centered at approximately 0.6 nm) and mesoscopic droplets (ranging from 30 to 500 nm in diameter). The persistence of the molecular aggregates across the entire pH range suggests their remarkable stability, implying that these nanostructures are highly resistant to disruption by ionic species present in the solution. Their robust stability indicates that molecular interactions such as hydrogen bonding and van der Waals forces contribute significantly to the self-assembly process, making them insensitive to the presence of external electrolytes.

In contrast, the mesoscopic droplets, whose size distribution ranges broadly between 30 nm and 500 nm, exhibit a more dynamic response to pH variation. Notably, the upper size limit of these droplets remains relatively consistent, fluctuating between 300 nm and 500 nm across most pH values, with the exception of pH 12, where larger droplets up to about 1 $\mu$m were observed. This suggests that highly alkaline conditions significantly influence the droplet coalescence process, potentially by altering interfacial tension and destabilizing the dispersed phase. In the near-neutral pH range (pH 6$\sim$8), the mesoscopic droplet sizes remain relatively stable, indicating an equilibrium state where intermolecular forces achieve a balance between repulsion and attraction.

Although the size distribution of the mesoscopic droplets does not exhibit significant variation across the pH range, the total scattering intensity rate (TSIR) shown in the inset of Figure~\ref{O15_dls}(a) provides additional insights into the structural changes within the system. The TSIR, which represents the cumulative scattering effect of all dispersed domains, exhibits a gradual decline with increasing pH, particularly above pH 10. This decrease in scattering intensity suggests a reduction in the number of dispersed droplets rather than a decrease in their individual sizes. This hypothesis is supported by NTA measurements, which can directly quantify droplet concentrations and validate the observed trends. Visualization techniques, as shown in Figure~\ref{O15_dls}(b) and Figure S2 in Supplementary Material, confirm that mesoscopic droplet nucleation is more prevalent in the mid-pH range, while an excess of alkaline solution at higher pH levels suppresses droplet formation, possibly by reducing interfacial charge and promoting coalescence. 

Further investigations into the statistical number density of mesoscopic droplets for two SFME formulations, including Sample A and Sample B, reveal consistent trends. As depicted in Figure~\ref{O5_dls}(a), mesoscopic droplets are most likely to nucleate and remain stable near neutral pH conditions. Extreme acidic (pH\textless4) or alkaline (pH\textgreater10) environments are less conducive to droplet formation, possibly due to charge screening effects and the destabilization of interfacial layers. Interestingly, the data indicate that the solution environment around pH 6 represents an optimal condition for mesoscopic droplet nucleation, suggesting that the interplay between molecular interactions and interfacial tension is most favorable under these conditions. In Sample B, where the aqueous phase content is lower, an increase in pH from 5 to 11 results in an 84\% decrease in mesoscopic droplet concentration. At pH 12, mesoscopic droplets are almost undetectable due to the significant reduction in scattering intensity, implying that the high ionic strength disrupts the delicate balance required for stable droplet formation.

The size-concentration distribution of mesoscopic droplets under different pH levels, illustrated in Figure~\ref{O5_dls}(b), further supports these findings. At neutral pH, the droplets exhibit a bimodal distribution, with a dominant peak centered around 300 nm. However, as acidity or alkalinity increases, the concentration of droplets declines, and the middle well-defined peak broadens and diminishes. This behavior suggests that strong acidic or basic conditions hinder the nucleation process by affecting interfacial energy barriers and molecular interactions~\cite{liu2001interfacial}. The characteristic droplet diameter $D_{50}$, representing the median size, shifts from 390 nm at pH 2 to 420 nm at pH 6 and further to 590 nm at pH 12, reflecting the tendency of droplets to become larger under extreme conditions due to reduced stabilization impact. DLS results on size distribution also show the similar trend as shown in Figure S3 in Supplementary Material. It is important to note that the characterization of nanobubbles and mesoscopic droplets by DLS and NTA techniques yields slightly different results. Unlike NTA, which directly counts particles, DLS measurements depend on light scattering intensity. As reported by Robert Botet et al.~\cite{botet2016interactions}, the mean droplet size is closely related to the relative rates of molecular diffusion and interface diffusion during mixing, indicating that the formation of nanostructures is highly dependent on the dynamic interplay of diffusion processes and external conditions such as electrolyte concentration and pH.

\begin{figure}[!t]
\centering
\includegraphics[width=0.45\textwidth]{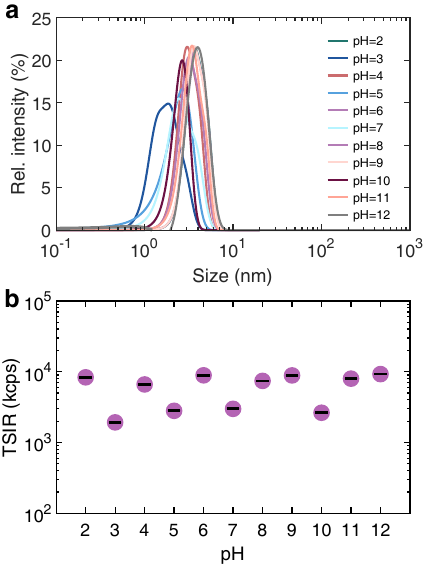}
\caption{(a) Size distribution of the reverse aggregates under various pH levels for Sample C with composition of 40 vol$\%$ trans-anothel, 55 vol$\%$ ethanol and 5 vol$\%$ water. (b) Total scattering intensity rate (TSIR) as a function of pH.} 
\label{O40}
\end{figure}

The nucleation behavior of inverse-SFME (W/O structuring) in response to solution pH presents unique characteristics that depend on the specific composition of the ternary system. SFME systems can exhibit diverse mesoscale structuring based on the relative proportions of trans-anethol/ethanol/water. In the case of Sample C, composed of 40 vol$\%$ trans-anethol, 55 vol$\%$ ethanol, and 5 vol$\%$ water, the pH-dependent structural responses were systematically analyzed using DLS, as illustrated in Figure~\ref{O40}(a). The DLS results indicate the presence of a well-defined monophasic region characterized by the formation of stable mesoscopic structures in the size range of 1 to 10 nm, with a dominant peak around 3 nm. Notably, this peak exhibits minimal shifts across the investigated pH range. This stability suggests that the structural organization of the system is highly resilient to pH fluctuations. The total scattering intensity rate (TSIR), presented in Figure~\ref{O40}(b), further supports this observation, showing a nearly constant intensity fluctuating between $10^3$ and $10^4$ kcps, indicating that the system's mesoscale organization is not significantly influenced by pH adjustments. The fluctuations observed in the data points are attributed to the inherent measurement procedure of the instrument (attenuator levels) rather than the variations in water aggregates present in the emulsion.

The observed stability can be primarily attributed to the low water content in Sample C, which minimizes the influence of pH regulation on the structuring process. In systems with higher water content, changes in pH can significantly alter the ionization of water molecules, leading to changes in the dielectric properties and aggregation behavior of mesoscopic droplets. However, in the current reverse-SFME system, the relatively hydrophobic environment limits the ability of ions to alter the interfacial properties of the droplets. The dielectric constant of the bulk phase remains relatively low. Here the stability of inverse-SFME structures is largely driven by molecular packing and solvation forces provided by ethanol and trans-anethol molecules. These components create a tightly packed interfacial layer that is relatively resistant to perturbations in the pH of the external environment. From a molecular configuration perspective, the formation of water-rich aggregates in the oil matrix suggests that significant charge transfer processes are unlikely to occur, as the micro-environment remains predominantly non-polar, thereby reducing ion mobility and interaction strength~\cite{han2022formation}. The low concentration of hydronium ($\rm{H_3O^+}$) or hydroxide ($\rm{OH^-}$) ions limits their ability to affect the interfacial charge or to penetrate deeply into the dispersed phase.

\begin{figure}[t!]
\centering
\includegraphics[width = 0.47\textwidth]{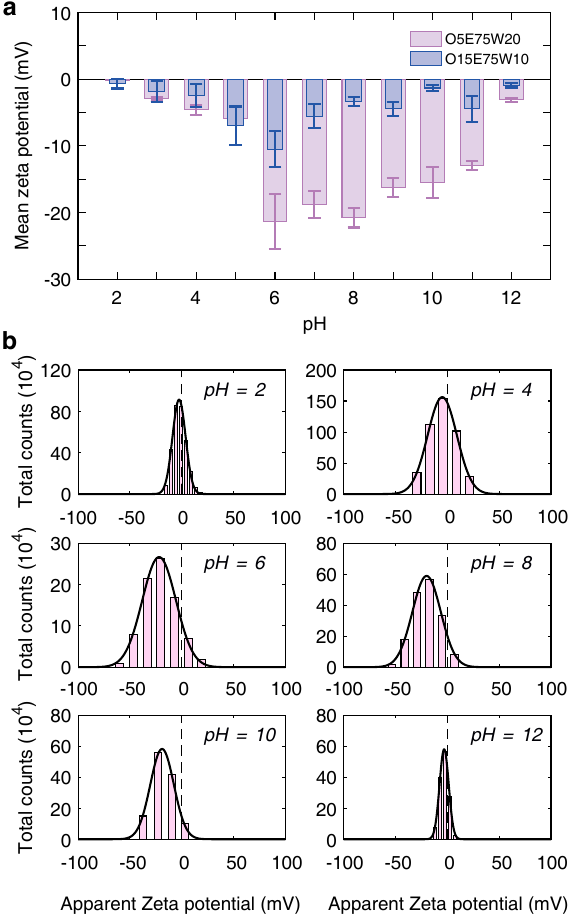}
\caption{(a) Mean Zeta potential as a function of the pH for two samples (Sample A with composition of 5 vol$\%$ trans-anothel, 75 vol$\%$ ethanol and 20 vol$\%$ water; Sample B with composition of 15 vol$\%$ trans-anothel, 75 vol$\%$ ethanol and 10 vol$\%$ water). (b) Zeta potential distribution under various pH levels for Sample A. } 
\label{O5_Zeta}
\end{figure}

One possible explanation for the pH-independent behavior of inverse-SFME structures lies in the difference between bulk and surface pH values, which arises due to variations in the degree of ionization and surface potential effects. To gain deeper insight into the surface charge characteristics of mesoscopic droplets, Zeta potential measurements were conducted using electrophoretic laser light scattering, with the results displayed in Figure~\ref{O5_Zeta}. Across the entire pH range, the mesoscopic droplets exhibit a negative surface charge, suggesting a consistent anionic nature of the droplet interface. The negatively charged properties of SFME droplets are also consistent with previous studies, although the magnitude is significantly different~\cite{rak2018mesoscale, shenavaei2020production, xu2013surfactant}. However, the magnitude of the Zeta potential varies with pH and differs between the two sample sets. In both cases, the mesoscopic droplets tend to approach electrical neutrality under highly acidic (pH 2) and alkaline (pH 12) conditions, with surface potential values converging to nearly zero, indicating reduced electrostatic interactions. 

The Zeta potential data reveal that the droplets exhibit their most negative charge near pH 6, where the measured values are -21.5 mV for Sample A and -10.5 mV for Sample B. This trend aligns with the observed size distribution data, highlights the idea that near-neutral conditions provide an optimal balance for mesoscopic droplet formation and stabilization. The difference in Zeta potential magnitude between the two samples can be attributed to their water content, with Sample A containing a higher proportion of water, which enhances the solvation of ions and facilitates their diffusion from the bulk phase to the droplet interface. Consequently, droplets in Sample A have stronger negative charges compared to those in Sample B.

Figure~\ref{O5_Zeta}(b) provides a detailed analysis of the surface potential distribution of the mesoscopic droplets. The data suggest that, in near-neutral pH conditions, negatively charged mesoscopic droplets dominate the population, whereas in strongly acidic or basic environments, the population becomes nearly electrically neutral, with a narrow potential distribution ranging between -10 mV and +10 mV. This observation implies that at extreme pH values, the system undergoes significant charge depletion, most likely leading to the destabilization and eventual collapse of mesoscopic droplets. The presence of both negatively and partially positively charged droplets suggests complex interfacial phenomena, potentially involving the preferential adsorption of cations or anions~\cite{lacour2025role}. In near-neutral conditions, the dominant negative charge indicates that the droplet interface favors the adsorption of hydroxide ions or negatively charged species. As the solution becomes more acidic or basic, competitive adsorption effects and charge screening mechanisms come into play, leading to a gradual neutralization of surface charges, reducing the stability of the droplets and potentially leading to phase transitions (close to the miscibility gap). 

The observed trends align with the findings reported by Robert Botet et al.~\cite{botet2016interactions}, which suggest that the size and stability of monodispersed emulsions are intricately linked to the balance between molecular diffusion and interfacial diffusion processes. In the present system, the limited water content constrains the influence of bulk pH, reinforcing the idea that interfacial phenomena dominate the stabilization and nucleation of mesoscopic structures. In addition, the abnormal response of mesoscopic droplets to alkaline regulation discovered in this work has not been fully investigated in detail. Based on previous studies~\cite{graciaa1995zeta, ho1999electrokinetic, gray2009explanation}, it was expected that mesoscopic droplets are benefit from a weakly alkaline environment, manifested by higher number concentration and surface potential. While this study found a near-neutral environment is optimal for the nucleation of mesoscopic droplets. This is a complex and hot topic involving the fundamental question: the origin of charge at oil-water interface. In our work, the highly alkaline environment of the water not only inhibits the nucleation of mesoscopic droplets but also affects the overall phase behavior. Moreover, revealing the arrangement of hydrogen bonds and the molecular structures at droplet interface is important to uncover the interfacial charging in SFME system.  Further advanced studies such as surface-sensitive spectroscopy, the SAXS (Small-angle X-ray Scattering) or SANS (Small-angle Neutron Scattering) and MD (Molecular Dynamics) simulation could provide additional insights into the specific mechanisms governing charge exchange at the interface of nanodomains in surfactant-free microemulsions.  

\subsection{Effect of salt on the nucleation of multiscale nanodomains}

\begin{figure}[!t]
\centering
\includegraphics[width=0.45\textwidth]{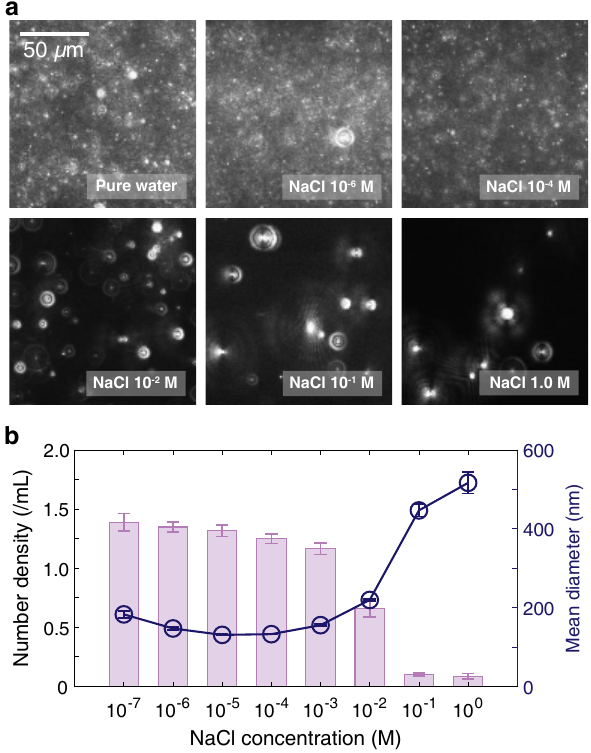}
\caption{Effect of NaCl on the formation of mesoscopic nanodroplets in SFME. (a) Snapshots of nanodroplets dispersed in ternary system. The concentration of salty ions ranged from 0 to 1.0 mol/L. (b) Number density and mean diameter of mesoscopic nanodroplets with the function of the NaCl concentration. The composition of the sample is 15 vol$\%$ trans-anothel, 80 vol$\%$ ethanol and 5 vol$\%$ water.} 
\label{salt}
\end{figure}

To further elucidate the role of added acid and base ions, such as hydroxide ($\rm{OH^-}$) and hydronium ($\rm{H_3O^+}$), we investigated the effects of sodium chloride (NaCl) on the self-assembly and stability of nanostructures within the SFME system. It is well known that salt ions, such as ($\rm{Na^+}$) and ($\rm{Cl^-}$), are generally considered hydrophilic, meaning their interaction with water molecules is characterized by high hydration enthalpies and entropies that significantly exceed the thermal energy scale ($kT$) of pure water. This strong affinity for water impacts the solution's microenvironment, leading to changes in phase behavior and nanostructure formation. Previous studies have demonstrated that, at sufficiently high salt concentrations, the addition of salts can reduce the mutual solubility of water and organic solvents, effectively shifting the system toward a more phase-separated state~\cite{marcus2015influence}. This phenomenon, commonly referred to as salting out, leads to the expansion of the two-phase region and destabilizes pre-Ouzo aggregates—weakly bound, transient nanostructures that are highly sensitive to electrolyte-induced perturbations.

Figure~\ref{salt}(a) presents NTA microscopic images, illustrating the nucleation of multi-scale nanodomains across different NaCl concentrations. The observed scattered light signal originates not only from the dispersed nanostructures but also from density fluctuations in the surrounding bulk liquid phase. Quantitative analysis, as shown in Figure~\ref{salt}(b), reveals that as the NaCl concentration surpasses $10^{-3}$ M, the number density of nucleated mesoscopic droplets decreases significantly. Specifically, nearly 90\% of the mesoscopic droplets disappear when the NaCl concentration reaches 1 M. Concurrently, the average droplet size increases from approximately 150 nm to 600 nm, suggesting a phase transition from mesoscopic pre-Ouzo aggregates to larger, well-defined Ouzo droplets, which marks the onset of phase demixing and enhanced oil-water separation.

We note that molecular dynamics simulations~\cite{schottl2018salt} have provided valuable insights into the behavior of ethanol and water at the oil-water interface in the presence of salts. These simulations indicate that salt ions drive ethanol molecules toward the interfacial region and into the oil-rich aggregates, effectively displacing ethanol from the bulk aqueous pseudo-phase. This phenomenon contributes to the salting-out effect, wherein the presence of salts reduces the water phase's solubilization capacity for organic components, thereby enhancing phase separation~\cite{ottosson2010influence}. By displacing ethanol from the aqueous pseudo-phase, the polarity of the water phase increases, shifting the system toward the two-phase region and leading to the formation of larger nanodroplets, as confirmed by experimental size histograms. The interaction of salt ions with water molecules results in significant enthalpic and entropic changes compared to pure water, which further promotes droplet coalescence and reduces the stability of smaller mesoscopic aggregates.

Ethanol's presence introduces additional complexity to the system. The hydroxyl group ($\rm{-OH}$) of ethanol molecules readily forms hydrogen bonds with water, creating a hydration shell that stabilizes the dispersed droplets. Ethanol, acting as a hydrotrope, reduces the oil-water interfacial tension, facilitating droplet dispersion and stability~\cite{liu2018surfactant}. However, short-chain alcohols, such as ethanol, do not form micelles but instead contribute to the formation of transient pre-Ouzo aggregates, which are typically around 140 nm in size and exhibit limited stability under high electrolyte concentrations. At elevated salt concentrations, most of the hydrophobic segments of trans-anethol-rich aggregates become screened by ethanol molecules, reducing the likelihood of ion adsorption at the oil-water interface. When ions do adsorb onto the aggregate surfaces, they preferentially localize in areas devoid of ethanol, which explains the diminished effect of salt ions on interfacial properties at lower concentrations~\cite{marcus2015influence}. 

The stability of SFMEs is highly influenced by electrolytes through mechanisms such as salting-out, differential ion adsorption, and interfacial stabilization. Anions can either promote phase separation by expelling ethanol into nanostructures or stabilize aggregates via electrostatic interactions, while cations typically increase aggregate size through salting-out. Antagonistic salts like $\rm NaBPh_4$ can convert pre-Ouzo aggregates into smaller and charged micelles~\cite{marcus2015influence}. MD simulations by Sebastian et al.~\cite{schottl2018salt}, further confirmed that salting-out of ethanol from aqueous phase and weak charging at the interface of octanol aggregates are dominant mechanisms in the presence of salts like NaI or LiCl addition. Additionally, ethanol at the interface diminishes the preferential binding effects of the ions. Although our macroscopic experimental data do not explicitly distinguish between different ion species, this does not imply that nanostructure nucleation and stability are entirely insensitive to ion-specific effects. Differences in ionic radius, hydration energy, and specific ion adsorption behavior could play a significant role in determining the final state of the system. Further investigations using molecular dynamics simulations are necessary to explore the spatial distribution of ions near the oil-water interface and to understand how specific ion properties influence SFME stability and phase behavior.

\subsection{Colloidal stability of mesoscopic droplets}

\begin{figure}[!t]
\centering
\includegraphics[width=0.47\textwidth]{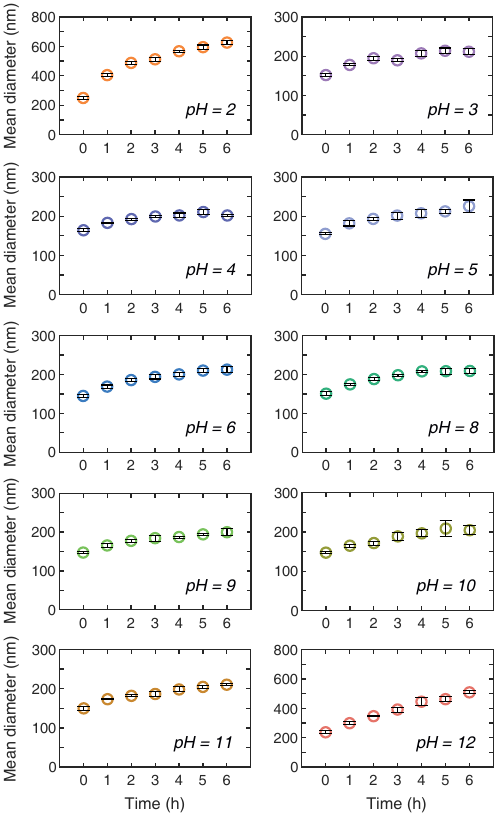}
\caption{Time dependence of the mean diameter of mesoscopic droplets under various pH for Sample A with composition of trans-anothel/ethanol/water 5/75/20 vol$\%$. The size of droplets was determined within six hours post-preparation. The error bar represents the standard deviation of the six measurements.} 
\label{O5_growth}
\end{figure}

The pH-dependent evolution of droplet size and growth velocity was systematically studied over a six-hour period. Growth of mesoscopic nanodroplets in sample A and sample B is shown in Figure~\ref{O5_growth} and Figure S4 in Supplementary Material, respectively. Although two cases have different mean diameter at certain time scale, the overall trends showing the droplet growth are similar. In general, mesoscopic droplets in this system do not behave as equilibrium entities; rather, they exhibit dynamic ripening processes over experimentally observable timescales. It was observed that under highly acidic conditions, droplet diameters increased from approximately 250 nm to 630 nm within six hours, exhibiting a unimodal growth profile. Conversely, under intermediate pH conditions (pH 3 to pH 11), droplet growth kinetics were significantly slower, with droplets expanding from 140 nm to 250 nm. The dynamic growth of nanodroplets in natural environment is governed by molecular diffusion, known as Ostwald ripening, rather than by aggregation-diffusion or coalescence. In this process, small nanodroplets do not come into direct contact, and the oil volumetric fraction in emulsions is relatively low. The driving force for growth is the difference in Laplace pressure between the droplets of different radii, leading to the enlargement of larger droplets at the expense of smaller ones. As reported by Natalia L. Sitnikova et al.~\cite{sitnikova2005spontaneously}, droplet growth is determined by the diffusive transport of dissolved matter through the dispersion medium rather than by coalescence. Herein, models such as collision-induced growth (CIG) model are not applicable for predicting the growth of nanodroplets. In strongly alkaline environments (pH 12), droplet growth accelerated, reaching sizes of 500 nm, which suggests that both diffusion-driven growth and coalescence contribute to size evolution under extreme pH conditions. 

The stability of mesoscopic droplets is intricately linked to the charge distribution at the oil-water interface. In emulsified systems, the oil-water interface can become enriched with ionic species, such as hydroxide ions ($\rm{OH^-}$), which play a critical role in stabilizing the droplets through the formation of an electric double layer. The differential sensitivity of direct and reverse nanostructures to ion concentration suggests that electrostatic forces within the double layer contribute significantly to droplet stability~\cite{tan2020bulk,zhang2020surface,doi2013soft}.

According to the double layer theory, a freshly formed droplet carries an initial surface charge density, denoted by $\sigma(R)$, which generates an outward-directed electrostatic pressure opposing the inward-directed Laplace pressure. The balance of these forces governs the droplet's mechanical stability. The electrostatic pressure exerted on the droplet surface is given by~\cite{tan2020bulk,zhang2020surface}
\begin{equation}
\label{eqn:PQ}
 p_{e}=\frac{\sigma^2}{2\epsilon\epsilon_0}
\end{equation}
where $\epsilon_0$ is the permittivity of free space, and $\epsilon$ represents the relative permittivity of the surrounding solution.

The presence of surface charges leads to the formation of a diffuse ionic cloud, which influences the overall electrostatic potential of the droplet. The surface charge density for a spherical droplet with radius can be described by
\begin{equation}
\label{eqn:sigmaR}
\sigma(R)=\frac{2\epsilon\epsilon_0k_BT}{\kappa^{-1}e}\sinh\left(\frac{e\psi}{2k_BT}\right)f(R)
\end{equation}
where $e=1.6\times10^{-19}$C is the elementary charge, $k_B$ is the Boltzmann constant, $T$ is temperature, $\psi$ is Zeta potential, $\kappa^{-1}=\sqrt{\frac{\epsilon\epsilon_0k_BT}{2Ie^2}}$ is the Debye length, $I$ is the ionic concentration. $f(R)$ is a geometric term arising from an approximate solution to the spherical Poisson-Boltzmann equation derived by Ohshima et al.~\cite{ohshima1983approximate,makino2010electrophoretic}, written as
\begin{equation}
\label{eqn:fR}
 f(R)=\sqrt{1+\frac{1}{\kappa R}\frac{2}{\cosh^2(\Psi/2)}+\frac{1}{(\kappa R)^2}\frac{8\ln[\cosh(\Psi/2)]}{\sinh^2(\Psi)}}
\end{equation}
where $\Psi=e\psi/2k_BT$. In the large droplet limit $\kappa R\gg 1$, $f(R)\xrightarrow{}1$, and Eq. (\ref{eqn:sigmaR}) becomes the Grahame equation for a planar double layer~\cite{israelachvili2011intermolecular,butt2018surface}. 

The growth of mesoscopic droplets is influenced by a delicate balance between electrostatic repulsion and interfacial tension. The latter is responsible for driving the system toward a thermodynamically favorable state by minimizing the interfacial energy. The equilibrium condition at the droplet interface can be expressed as:
\begin{equation}
\label{eqn:P}
P_{in}+P_e=P_{out}+\frac{2\gamma}{R}
\end{equation}
where $P_{in}$ is the pressure inside the droplet, $P_{out}$ is pressure of bulk phase solution around the droplet, $P_e$ is the electrostatic force of net charge loaded on the surface, ${2\gamma}/{R}$ is the Laplace pressure, and $\gamma$ is the surface tension of oil-water interface without charge adsorption. Ignoring the effect of charge adsorption at the droplet interface, Eq. (\ref{eqn:P}) can be derived from the classical laplace equation $\Delta P=P_{in}-P_{out}={2\gamma}/{R}$. The presence of electrostatic forces effectively reduces the Laplace pressure, resulting in slower ripening.

\begin{figure}[!t]
\centering
\includegraphics[width=0.46\textwidth]{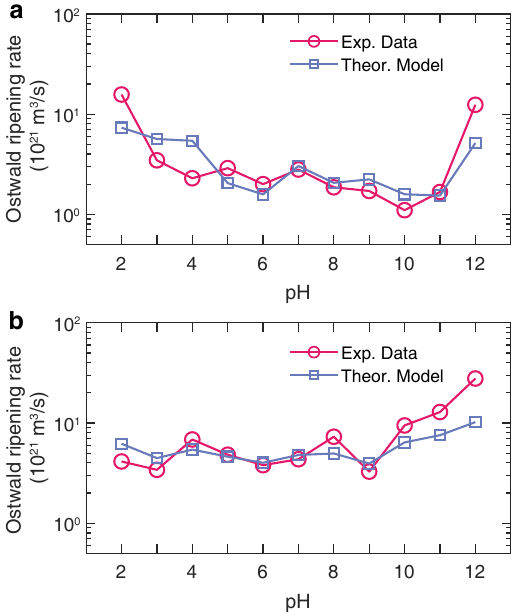}
\caption{Comparison of experimental and theoretically calculated (Eq.(\ref{v})) Ostwald ripening rates of mesoscopic droplets across the whole pH range for (a) Sample A with composition of trans-anothel/ethanol/water 5/75/20 vol$\%$ and (b) Sample B with composition of trans-anothel/ethanol/water 15/75/10 vol$\%$. } 
\label{theoritical}
\end{figure}

The primary mechanism driving droplet growth in SFME systems is Ostwald ripening, a process where smaller droplets dissolve and larger droplets grow due to differences in oil solubility driven by Laplace pressure gradients. The rate of droplet growth due to Ostwald ripening can be approximated by:
\begin{equation}
\label{eqn:v1}
v=\frac{dR^3}{dt}=\frac{4}{9}\frac{2\phi\gamma}{\bar{R}T}DC_{oil}^{sat}
\end{equation}
where $R$ is the average radius of the droplets, $D$ is the dispersed-phase molecular diffusion coefficient in the continuous phase, $C_{oil}^{sat}$is the aqueous solubility of the dispersed phase, $\phi$ is the molar volume of the dispersed phase, $\gamma$ is the interfacial tension between the dispersed and the continuous phases, and $\bar{R}$ and $T$ are the gas constant and the absolute temperature~\cite{sitnikova2005spontaneously}. 

Charge adsorption at the droplet interface modifies the effective interfacial tension, resulting in a corrected growth rate given by:
\begin{equation}
    v=\frac{dR^3}{dt}=\frac{4}{9}\frac{2\phi\gamma_{eff}}{\bar{R}T}DC_{oil}^{sat}
\label{v}
\end{equation}
where the effective interfacial tension $\gamma_{eff}$ is given by $2\gamma_{eff}=\Delta PR=(\frac{2\gamma}{R}-P_e)R$. This indicates that charge adsorption leads to a reduction in effective interfacial tension, slowing down the growth process and enhancing droplet stability.

The ripening rates of mesoscopic droplets in Sample A and B were analyzed by plotting the cube of the average droplet radius as a function of time. The slopes of these plots were compared with theoretical predictions based on the charge-stabilization model, revealing a reasonable agreement within a factor of three, as illustrated in Figure~\ref{theoritical}. This level of consistency suggests that the theoretical framework captures the essential physical mechanisms governing the droplet evolution, albeit with some deviations at extreme pH values. The ability of this theoretical model to approximate experimental results across a wide pH range, particularly under highly acidic and alkaline conditions, further supports the hypothesis that Ostwald ripening is the dominant aging mechanism of these mesoscopic droplets. 

Notably, at the pH extremes (below pH 3 and above pH 10), the ripening rates were significantly higher compared to intermediate pH conditions (pH 3$\sim$10). This accelerated growth at extreme pH levels suggests that the interfacial properties of the droplets are highly sensitive to ionic strength, which could facilitate molecular diffusion across the droplet interface and promote coalescence processes. In contrast, within the intermediate pH range, the slower growth rates indicate a relatively stable droplet structure, where electrostatic stabilization and steric hindrance provide effective resistance to Ostwald ripening.

Quantitative analysis of the ripening rates revealed values ranging from $10^{-21}$ to $\rm{3\times10^{-20}\ m^3/s}$, which are comparable to or slightly higher than those reported for both surfactant-free and surfactant-stabilized emulsions~\cite{taylor1998ostwald}. This suggests that our ternary system exhibits a ripening behavior consistent with other ternary systems. However, it is important to note that the ripening rate of these mesoscopic droplets was found to be at least two orders of magnitude higher than that of classical Ouzo droplets~\cite{sitnikova2005spontaneously}. This significant difference implies that the solubilization and interfacial dynamics of SFME systems are more favorable for rapid molecular exchange compared to traditional Ouzo-type dispersions, possibly due to the presence of ethanol as a cosolvent, which facilitates faster molecular diffusion.

Comparing the behavior of Sample A and Sample B, it was observed that Sample B exhibited a consistently higher ripening rate. This difference is likely attributed to the weaker surface potential of droplets in Sample B, which results in reduced electrostatic repulsion and a higher tendency for molecular exchange and coalescence. The data presented in Figure~\ref{theoritical}(b) indicate that for Sample B, ripening rates remained relatively unchanged when the pH did not exceed 9, suggesting that moderate electrolyte concentrations had little impact on the stability and growth of the droplets. However, at strongly alkaline conditions (pH $>$ 10), substantial deviations were observed between experimental data and theoretical predictions. This discrepancy may arise from complexities not fully captured by the existing model, such as pH-dependent changes in interfacial tension, droplet polydispersity, and the impact of ion-specific effects on droplet dynamics.

\section{Conclusions}

We systematically investigated the nucleation and growth behavior of surfactant-free microemulsions (SFMEs) comprising trans-anethol, ethanol and water under varying ionic conditions. As the pH ranged from 2 to 12, a U-shaped dependence of dielectric constant and conductivity was observed, indicating enhanced-ionic activity at acidic (pH \textless 3) and alkaline (pH \textgreater 10) extremes. Meanwhile, zeta potential measurements reveal that droplets exhibit the most negative surface charge in the neutral environment (pH 6–7), promoting stronger electrostatic repulsion with higher droplet number density. As the pH approached to the extreme, surface charges trended toward zero, facilitating droplet coalescence and phase separation. The heightened ionic strength at extreme pH levels (pH \textless 3 and pH \textgreater 10) destabilizes the dispersed phase. Furthermore, the effect of electrolyte (e.g., NaCl) on SFMEs was examined, showing a “salting-out” phenomenon and accelerating phase separation as salt concentration above $10^{-3}$M. Elevated salt level increased the interfacial tension and weakened electrostatic repulsion, leading to the formation of large droplets with lower number density.

Moreover, we conducted theoretical calculations to verify the primary mechanism of droplet growth. Both the experimental data and charge-stabilization theory confirmed the occurrence of Ostwald ripening, accurately describing the nanodroplet growth across the entire pH range. It should be noted that our theoretical calculations are derived from the stable equilibrium model of nanobubbles, which does not consider the effect of additives on surface tension. Future work will combine MD simulations to analyze the molecular structure configuration of interfaces to improve the feasibility of this model. Nevertheless, the model successfully predicts the growth rate of nanodroplets across a range of ionic environments. Specifically, at extreme pH levels, ripening accelerates due to reduced electrostatic repulsion and enhanced coalescence, while near-neutral pH conditions result in a slower ripening rate. 

These findings underscore the critical influence of pH on SFME stability and highlight opportunities to optimize formulation by tuning pH, electrolyte levels, and composition. Such insights may provide the guidance for the development of biocompatible drug carriers, advanced food emulsions and other eco-friendly industrial processes.

\section*{Declaration of Competing Interest}

The authors declare that they have no known competing financial interests or personal relationships that could have appeared to influence the work reported in this paper.

\section*{Acknowledgments}

This work is supported by the National Natural Science Foundation of China under Grants Nos. 11988102 and 12202244, New Cornerstone Investigator Program, the Xplorer Prize, SINOPEC Petroleum Exploration and Production Research Institute and Shui Mu Scholarship at Tsinghua University (2024SM049).

\appendix
\section*{Appendix A. Supplementary Material}

Supplementary data associated with this article can be found in a separate file. 

\bibliographystyle{elsarticle-num}
\bibliography{references}


\end{sloppypar}

\end{document}